\UseRawInputEncoding
\documentclass[10pt,a4paper]{article}
\usepackage{jheppub_kim}
\usepackage{pdflscape}
\usepackage{amsmath}
\usepackage{amssymb}
\usepackage{dcolumn}
\usepackage{bm}
\usepackage{color}
\usepackage{epsfig}
\usepackage{amsfonts}
\usepackage{graphicx}
\usepackage{subfigure}
\usepackage{dcolumn}
\usepackage{multirow}

\begin{document}
\title{Observational Constraint in $f(R,T)$ gravity from the cosmic chronometers and some standard distance measurement parameters}

\author[a]{Prabir Rudra,}
\author[b]{Kinsuk Giri}

\affiliation[a] {Department of Mathematics, Asutosh College,
Kolkata-700 026, India.}

\affiliation[b]{Department of CSE, National Institute of Technical
Teacher's Training and Research, Block-FC, Sector-III, Salt Lake,
Kolkata-700 106, India}

\emailAdd{prudra.math@gmail.com, rudra@associates.iucaa.in}
\emailAdd{kinsuk@nitttrkol.ac.in, kinsuk84@gmail.com}

\abstract{In this work we perform an observational data analysis
on the $f(R,T)$ gravity with the aim of constraining the parameter
space of the model. Five different models are considered and the
30 point $z-H(z)$ cosmic chronometer data is used in our analysis.
The $\chi^2$ statistic is formulated as a difference between the
theoretical and the observed values of $H(z)$ for different values
of redshift $z$. Efforts have been made to minimize the statistic
in order to get the best fit values for the free parameters
models. We have also used some standard distance measurement
parameters like BAO and CMB peaks along with the data for
achieving better constraints on the parameter space. We have used
the publicly available \textit{CosmoMC} code for obtaining the
bounds for the free parameters of the models in different
confidence intervals like $68\%$, $95\%$, $99\%$. 1D distributions
and 2D joint confidence contours for various confidence levels
mentioned above are generated for the free parameters using the
\textit{CosmoMC} code. Finally our models have been compared with
the standard $\Lambda CDM$ model by some statistical techniques
using the observational data and the support for the models from
the data is probed.}


\maketitle

\section{Introduction}
One of the major challenges of modern cosmology is to find a
suitable model of the universe that can incorporate the late
cosmic acceleration (Riess et al. 1998; Perlmutter et al. 1999;
Spergel et al. 2003). It has been over two decades that we are
involved in this quest, but still our wish has not been granted
and the target keeps eluding us. In principle general relativity
(GR) is the best option that we have as a candidate of a theory of
gravity and we have almost subscribed to this fact unanimously.
But it is seen that it degenerates at cosmological distances and
does not have any provision to explain the accelerated expansion
of the universe. This has prompted us to look for other options
where we may be able to theorize this observed phenomenon. With
more and more people bent on doing this various ideas began to
flourish in literature. After careful scrutiny, all these ideas
may be categorized into two types, which are dark energy (DE) and
modification of Einstein's gravity. Since GR is a theory that
connects the matter content of the universe with the curvature of
space time, it is obvious that any sort of modification must be
carried out on one of these two components. Dark energy is a
concept via which the matter content of the universe is modified
from usual matter to an exotic fluid which has an anti-gravity
effect, which may be employed to explain the cosmic acceleration.
The reader is encouraged to consult the ref. Brax (2018) for
extensive reviews on DE.

Since GR is a geometric theory of gravitation, it would be logical
to think that by introducing new forms of geometrical structure,
we will be able to modify the standard geometry used in GR and
such modifications may prove to be fruitful in our probe for an
alternative model. This is the theory of modified gravity.
Detailed and comprehensive reviews in modified gravity can be
found in the refs. (Nojiri, Odintsov \& Oikonomou 2017; Nojiri \&
Odintsov 2007). It is known that the field equations of GR are
derived from an action principle using the Einstein-Hilbert (EH)
action. In EH action the gravity Lagrangian is given by the Ricci
scalar invariant $R$. The most obvious modification to this is
brought about by replacing the Ricci scalar, $R$ in EH action by
an analytic function $f(R)$. This is the $f(R)$ gravity theory.
Considering various forms of $f(R)$ models, one can explore the
different forms of non-linear effects of the scalar curvature. The
reader may refer to the refs. (De Felice \& Tsujikawa 2010 and
Sotiriou \& Faraoni 2010) where detailed reviews on $f(R)$
theories have been provided. Although mathematically any form of
$f(R)$ model is allowed, but there are some models which may be
ruled out because of their non-agreement with cosmological
observations. Such models were discussed by Amendola, Polarski \&
Tsujikawa (2007). A cosmological dynamical system analysis in
$f(R)$ gravity was perfomed by Amendola et al. (2007). A study of
large scale structure in the background of $f(R)$ gravity was
conducted by Song, Hu \& Sawicki (2007). Nojiri \& Odintsov (2006)
studied various reconstruction schemes using $f(R)$ gravity.

In due course it was understood that a dynamical relationship
between matter and curvature of spacetime can produce some very
interesting models which can solve many existing problems of
cosmology. This idea gave rise to coupling models between matter
and geometry. Two different forms of coupling, namely minimal and
non-minimal coupling (NMC)  (Azizi \& Yaraie 2014) between
geometry and matter content was established and studied
extensively. Particularly NMC theories are very useful in solving
various problems like, providing explanation for post inflationary
pre heating (Bertolami et al., 2011), large scale structure
formation (Nesseris 2009; Bertolami et. al 2013; Thakur \& Sen
2013 ), etc. They were also utilized to mimic dark energy
(Bertolami et al. 2010, Bertolami \& Pramos 2011) and dark matter
(Bertolami \& Pramos 2010, Harko 2010). The concept of NMC is used
on a large scale to couple geometry with matter in the form of
scalar fields (Futamase \& Maeda 1989; Fakir \& Unruh 1990; Uzan
1999; Amendola 1999; Torres 2002) giving rise to scalar tensor
theories. In this connection a particular class of theories known
as $f(R,L_{m})$ theories was proposed by Harko \& Lobo (2010),
where $\L_{m}$ represents the matter Lagrangian. Further
developments in this theory can be found in Azevedo \& Pramos
(2016) and Pourhassan \& Rudra (2020). A specific interesting sub
class of these theories was proposed by Harko et al. (2011), where
the authors represented the matter Lagrangian by the trace $T$ of
the energy-momentum tensor (EMT) $T_{\mu\nu}$. This theory is
known as the $f(R,T)$ theory of gravity, where the gravitational
Lagrangian is an analytic function of two scalar invariants $R$
and $T$. It is seen from the derivation of the field equations of
this theory that they depends upon a source term, which in turn
depends upon the variation of the EMT with the metric. So, it is
quite clear that the form of the field equations will completely
depend upon the nature of matter content of the universe. The
association between matter and geometry can be described via the
function $f(R,T)$ is different ways. These options basically
involve both minimal and non-minimal coupling between matter and
curvature. Moreover we see that the covariant divergence of the
EMT in non-vanishing for this theory, which results in
non-geodesic motion for the massive test particles. This is
attributed to the additional acceleration on the particles due to
the coupling effects of matter and geometry. This is an
interesting feature of the theory, which draws a lot of attention.
Probably this is why we have seen considerable developments to
this theory over the years in the literature. A thermodynamic
study in the background of $f(R,T)$ gravity was performed by
Sharif and \& Zubair (2012). The problem of cosmic coincidence in
this theory was studied in Rudra (2015). Cosmological phase space
analysis in $f(R,T)$ theory was performed by Shabani \& Farhoudi
(2013). Scalar perturbations in $f(R,T)$ gravity were explored by
Alvarenga et al. (2013). Gravastars in the background of $f(R,T)$
framework was studied by Das et al. (2017). Zaregonbadi et al.
(2016) explored the dark matter effects arising out of the
$f(R,T)$ models. Polar gravitational waves and their evolution was
studied by Sharif \& Siddiqa (2019). Gravitational collapse of
$f(R,T)$ models in Vaidya spacetime along with cosmic censorship
hypothesis was studied in Rudra (2020).

 No matter how promising a theory may seem to be, it has to comply with the observations in order to establish itself as a acceptable physical theory. In cosmology observational data analysis of physical models helps us to check the viability between theoretical and observational results. In most of these studies we use statistical techniques as tools to reconcile data with theory. We know that all theoretical models possess various free parameters. By fitting these theoretical models with the observational data-sets (retrieved from various space probes) it is possible to constrain the free parameters using suitable statistical tests. Once the values of the free model parameters are determined, the model becomes self sufficient and deterministic in nature, which may then be used to probe other cosmological issues. In this work we are motivated to perform such an analysis on $f(R,T)$ theory. The idea is to consider some generic $f(R,T)$ models and try to constrain their parameter space with observational data. The motivation for this work is very high, since such an analysis will help us to identify some cosmological viable models which can be further used to verify other cosmological issues. The paper is organized as follows: In section II we have reviewed the basic equations of $f(R,T)$ gravity. In section III, we have performed a detailed observational data analysis of some models using cosmic chronometer data and some standard distance measurement parameters like BAO and CMB peaks. Section IV is dedicated to model comparison using some statistical criterion and finally the paper ends with a discussion and conclusion in section V.

\section{Basic equations of $f(R,T)$ gravity}
The Einstein-Hilbert action for general relativity is given by,
\begin{equation}\label{actionEH}
S_{EH}=\frac{1}{2\kappa}\int R\sqrt{-g}d^{4}x
\end{equation}
where $\kappa\equiv 8\pi$, $g$ is the determinant of the metric
and $R$ is the Ricci scalar (we have considered $G$=$c=1$). We
replace the Ricci scalar, $R$ in the above action by a generalized
function of $R$ to get the action for $f(R)$ gravity (Sotiriou et
al. 2010, de Felice et al. 2010),
\begin{equation}\label{actionR}
S=\frac{1}{2\kappa}\int f(R)\sqrt{-g}d^{4}x
\end{equation}
Taking the action (\ref{actionR}) and adding a matter term $S_M$,
the total action for $f(R)$ gravity takes the form,
\begin{equation}\label{actionRtotal}
S_{f(R)}=\frac{1}{2\kappa}\int f(R)\sqrt{-g}d^{4}x+\int
\mathcal{L}_{m}\sqrt{-g}d^{4}x
\end{equation}
where $\mathcal{L}_m$ is the matter Lagrangian and the second
integral on the R.H.S is $S_M$ representing the matter fields. To
obtain the action for $f(R,T)$ gravity we further modify the
action for $f(R)$ gravity by introducing the trace of the
energy-momentum tensor $T_{\mu\nu}$ in the gravity Lagrangian as
follows (Harko et al. 2011),
\begin{equation}\label{action}
S_{f(R,T)}=\frac{1}{2\kappa}\int f(R,T)\sqrt{-g}d^{4}x+\int
\mathcal{L}_{m}\sqrt{-g}d^{4}x+\int
\mathcal{L}_{rad}\sqrt{-g}d^{4}x
\end{equation}
Here $f(R,T)$ is an arbitrary function of the Ricci scalar $R$ and
the trace $T$ of the energy-momentum tensor $T_{\mu\nu}$. Here we
have considered radiation along with matter as a component of the
universe, which is represented by the action integral for
radiation on the RHS. The energy-momentum tensor is defined as
(Landau \& Lifshitz 1998),
\begin{equation}\label{energymomentum}
T_{\mu\nu}=-\frac{2}{\sqrt{-g}}\frac{\delta
(\sqrt{-g}\mathcal{L}_{m})}{\delta g^{\mu\nu}}
\end{equation}
The trace of this tensor can be given as $T=g^{\mu\nu}T_{\mu\nu}$.
Taking variation with respect to the metric we get the field
equations for $f(R,T)$ gravity as,
\begin{equation}\label{field}
f_{R}(R,T)R_{\mu\nu}-\frac{1}{2}f(R,T)g_{\mu\nu}+\left(g_{\mu\nu}\Box-\nabla_{\mu}\nabla_{\nu}\right)f_{R}(R,T)=\kappa
T_{\mu\nu}-f_{T}(R,T)T_{\mu\nu}-f_{T}(R,T)\Theta_{\mu\nu}+\kappa
T_{\mu\nu}^{rad}
\end{equation}
where $\Theta_{\mu\nu}$ is given by,
\begin{equation}\label{stressenergy}
\Theta_{\mu\nu}\equiv g^{\alpha\beta}\frac{\delta
T_{\alpha\beta}}{\delta g^{\mu\nu}}
\end{equation}
In the field equations $\nabla_{\mu}$ denotes covariant derivative
associated with the Levi-Civita connection of the metric and
$\Box\equiv \nabla^{\mu}\nabla_{\mu}$ is the D'Alembertian
operator. Moreover we have denoted $f_{R}(R,T)=\partial
f(R,T)/\partial R$ and $f_{T}(R,T)=\partial f(R,T)/\partial T$.
The tensor $\Theta_{\mu\nu}$ can be calculated as,
\begin{equation}\label{theta}
\Theta_{\mu\nu}=-2T_{\mu\nu}+g_{\mu\nu}\mathcal{L}_{m}-2g^{\alpha\beta}\frac{\partial^{2}\mathcal{L}_{m}}{\partial
g^{\mu\nu}\partial g^{\alpha\beta}}
\end{equation}
It is seen that the above tensor depends on the matter Lagrangian.
For perfect fluid the above tensor becomes,
\begin{equation}\label{thetaperfect}
\Theta_{\mu\nu}=-2T_{\mu\nu}+pg_{\mu\nu}
\end{equation}
If we consider pressure-less dust, then $p=0$ and using
eq.\eqref{thetaperfect} in eq.\eqref{field} we get,
\begin{equation}\label{field2}
f_{R}(R,T)R_{\mu\nu}-\frac{1}{2}f(R,T)g_{\mu\nu}+\left(g_{\mu\nu}\Box-\nabla_{\mu}\nabla_{\nu}\right)f_{R}(R,T)=\left(\kappa++f_{T}(R,T)\right)
T_{\mu\nu}+\kappa T_{\mu\nu}^{rad}
\end{equation}
Contracting the above equation we get,
\begin{equation}\label{tracefield}
Rf_{R}(R,T)+3\Box
f_{R}(R,T)-2f(R,T)=\left(\kappa+f_{T}(R,T)\right)T
\end{equation}

Now we consider a spatially flat
Friedmann-Lemaitre-Robertson-Walker (FLRW) spacetime,
\begin{equation}\label{frwmetric}
ds^{2}=-dt^{2}+a^{2}(t)\left(dx^{2}+dy^{2}+dz^{2}\right)
\end{equation}
where $a(t)$ is the cosmological scale factor. Using
eqns.\eqref{field2}, \eqref{tracefield} and \eqref{frwmetric} we
get the following FLRW equations,
\begin{equation}\label{frw1}
3H^{2}f_{R}(R,T)+\frac{1}{2}\left(f(R,T)-Rf_{R}(R,T)\right)+3H\frac{d}{dt}f_{R}(R,T)=\left(\kappa+f_{T}(R,T)\right)\rho_{m}+\kappa
\rho_{rad}
\end{equation}
and
\begin{equation}\label{frw2}
2\dot{H}f_{R}(R,T)+\frac{d^{2}}{dt^{2}}f_{R}(R,T)-H\frac{d}{dt}f_{R}(R,T)=-\left(\kappa+f_{T}(R,T)\right)\rho_{m}-\frac{4}{3}\kappa
\rho_{rad}
\end{equation}
where $H=\frac{\dot{a}(t)}{a(t)}$ is the Hubble parameter and
$\rho_{m}$, $\rho_{rad}$ are the energy densities of matter and
radiation respectively. The above equations may be written in the
form of the standard FLRW equations as,
\begin{equation}
3H^{2}=\kappa
\rho_{eff}=\kappa\left(\rho_{m}+\rho_{mod}+\rho_{rad}\right)
\end{equation}
and
\begin{equation}
2\dot{H}+3H^{2}=-\kappa\left(\rho_{eff}+p_{eff}\right)=-\kappa\left(\rho_{m}+\rho_{mod}+\rho_{rad}+p_{mod}\right)
\end{equation}
where
\begin{equation}
\rho_{mod}=\frac{-f(R,T)-6H\frac{d}{dt}f_{R}(R,T)+2\left(\kappa+f_{T}(R,T)\right)\rho_{m}+2\kappa
\rho_{rad}+f_{R}(R,T)\left(R-2\kappa\left(\rho_{m}+\rho_{rad}\right)\right)}{2\kappa
f_{R}(R,T)}
\end{equation}
and
\begin{equation}
p_{mod}=\frac{3\left[\frac{d^{2}}{dt^{2}}f_{R}(R,T)+f(R,T)+5H\frac{d}{dt}f_{R}(R,T)-Rf_{R}-\left(f_{T}(R,T)+\kappa\right)\rho_{m}\right]-2\kappa
\rho_{rad}}{3\kappa f_{R}(R,T)}
\end{equation}
Here $\rho_{mod}$ and $p_{mod}$ are the energy density and
pressure contributions respectively from the modified gravity.
These can be considered equivalent to the contributions from a
dark fluid component. Moreover $\rho_{eff}$ and $p_{eff}$ are
respectively the effective energy density and pressure of the
model.

Moreover the continuity equations for various components are given
as follows,
\begin{equation}\label{matcons}
\dot{\rho}_{m}+3H\rho_{m}=0
\end{equation}

\begin{equation}\label{decons}
\dot{\rho}_{mod}+3H\left(\rho_{mod}+p_{mod}\right)=0
\end{equation}

\begin{equation}\label{radcons}
\dot{\rho}_{rad}+4H\rho_{rad}=0
\end{equation}
Here we have neglected the pressures of matter and radiation
components. Solving eqns.\eqref{matcons} and \eqref{radcons} we
get the energy densities of matter and radiation components
respectively as $\rho_{m}=\rho_{m0} \left(1+z\right)^{3}$ and
$\rho_{rad}=\rho_{rad0} \left(1+z\right)^{4}$. Here $\rho_{m0}>0$
and $\rho_{rad0}>0$ are constants that represents the current
energy densities of matter and radiation respectively. Moreover
$z=\frac{1}{a(t)}-1$ represents the cosmological redshift. In
general we have the energy momentum scalar invariant
$T=\rho_{m}+3p_{m}$. But since here we have considered
pressure-less dust, the expression becomes $T=\rho_{m}$. Now in
order to give it a generic nature here we will consider $T=\lambda
\rho_{m}$, where $\lambda$ is a constant. The effective equation
of state parameter can be given by,
\begin{equation}\label{eoseff}
w_{eff}=\frac{p_{eff}}{\rho_{eff}}=\frac{p_{mod}}{\rho_{m}+\rho_{mod}+\rho_{rad}}
\end{equation}

\section{Observational data analysis}
Here, we would like to perform observational data analysis on our
theoretical model using observational data-sets. In order to
perform this rigorous analysis, we will use different statistical
techniques, viz., $\chi^2$ minimization techniques, Markov chain
Monte Carlo random sampling methods, etc. We have written codes in
PYTHON for $\chi^2$ minimization techniques, while visualizations
and plotting are done by the publicly available \textit{CosmoMC}
code (Lewis et al. 2000; Lewis \& Bridle 2002). We would also like
to check the validity of our theoretical models by the amount of
support they get from the observational data. Below we start our
analysis by considering the models. Then we will consider our
data-set that we will use in this study. Finally we will perform
the data fitting analysis. We will also consider some distance
measurement parameters like BAO and CMB peaks in our analysis to
further constrain the models. This will reduce the degeneracy
between the free parameters of the models.

\subsection{The Models}
Here we will discuss the different $f(R,T)$ models that we will
use in our study. Henceforth in all the models we will consider
$\kappa=1$. In our models we will consider both minimal and
non-minimal coupling between the matter and curvature.

\subsubsection{Model I: Minimal coupling in power law form}
The model is given by,
\begin{equation}
f(R,T)=\alpha R^{n}+\beta T^{m}
\end{equation}
where $\alpha$, $\beta$, $n$ and $m$ are constant parameters.

In this model for $n=1$, the first FLRW equation \eqref{frw1}
becomes,
\begin{equation}\label{mod11}
H^{2}(z)=\frac{1}{3\alpha}\left[\left\{\rho_{m0}+\rho_{rad0}\left(1+z\right)\right\}\left(1+z\right)^{3}-\frac{1}{2\lambda}\left\{\beta\left(\lambda-2m\right)\left(\lambda
\rho_{m0}\left(1+z\right)^{3}\right)^{m}\right\}\right]
\end{equation}
Now we will define the dimensionless density parameters as,
\begin{equation}\label{denpara}
\Omega_{m0}=\frac{\rho_{m0}}{3H_{0}^{2}},~~~~~~~~\Omega_{rad0}=\frac{\rho_{rad0}}{3H_{0}^{2}}
\end{equation}
We define another dimensionless parameter for expansion rate as,
\begin{equation}\label{dimexp}
E(z)\equiv \frac{H(z)}{H_{0}}
\end{equation}
Using these parameters eqn.\eqref{mod11} can be written as,
\begin{equation}\label{mod111}
E(z)=\frac{1}{\sqrt{\alpha}}\left[\left\{\Omega_{m0}+\Omega_{rad0}\left(1+z\right)\right\}\left(1+z\right)^{3}-\frac{1}{2\lambda}\left\{\beta\left(\lambda-2m\right)\left(3H_{0}^{2}\right)^{m-1}\left(\lambda
\Omega_{m0}\left(1+z\right)^{3}\right)^{m}\right\}\right]^{1/2}
\end{equation}
Here we see that the free parameters appearing in the model are
$H_{0}$, $\Omega_{m0}$, $\Omega_{rad0}$, $\lambda$, $\alpha$,
$\beta$ and $m$. We will fix some of these parameters using the
best-fit values from 7-year WMAP data (Komatsu et al, 2011). We
fix the parameters $H_{0}$, $\Omega_{m0}$, $\Omega_{rad0}$ by
considering the values $H_{0}=72 Km/sec/Mpc$, $\Omega_{m0}=0.3$
and $\Omega_{rad0}=10^{-4}$. So we are left with only four free
parameters in this model and the corresponding parameter space to
be constrained is $\left(\lambda, \alpha, \beta, m\right)$.

\subsubsection{Model II: Minimal coupling in exponential form}
The model is given by,
\begin{equation}\label{mod2}
f(R,T)=\alpha R+\beta e^{m T}
\end{equation}
where $\alpha$, $\beta$ and $m$ are constant parameters. For this
model the first FLRW equation becomes,
\begin{equation}\label{mod22}
H^{2}(z)=\frac{1}{6\alpha}\left[2\left\{\rho_{m0}+\rho_{rad0}\left(1+z\right)\right\}\left(1+z\right)^{3}+\beta\left(2m\rho_{m0}\left(1+z\right)^{3}-1\right)e^{m\lambda
\rho_{m0}\left(1+z\right)^{3}}\right]
\end{equation}
Using the dimensionless parameters defined in the
eqns.\eqref{denpara} and \eqref{dimexp} we get from the above
equation,
\begin{equation}\label{mod222}
E(z)=\frac{1}{\sqrt{2\alpha}}\left[2\left\{\Omega_{m0}+\Omega_{rad0}\left(1+z\right)\right\}\left(1+z\right)^{3}+\beta\left(2m\Omega_{m0}\left(1+z\right)^{3}-\left(3H_{0}^{2}\right)^{-1}\right)e^{3H_{0}^{2}m\lambda
\Omega_{m0}\left(1+z\right)^{3}}\right]^{1/2}
\end{equation}
Similar to previous model, here the working parameter space to be
constrained is $\left(\lambda, \alpha, \beta, m\right)$.

\subsubsection{Model III: Pure Non-minimal coupling}
The model is given by,
\begin{equation}\label{mod3}
f(R,T)=f_{0}R^{n}T^{m}
\end{equation}
where $f_{0}\neq 0$, $n$ and $m$ are constant parameters. For this
model using the FLRW equations we get for $n=1$,
\begin{equation}\label{mod33}
H^{2}(z)=\frac{\left[\lambda\rho_{m0}+\rho_{rad0}\left(\lambda-m\right)\left(1+z\right)\right]\left(1+z\right)^{3}}{3f_{0}\left(\lambda-m\right)\left(\lambda\rho_{m0}\left(1+z\right)^{3}\right)^{m}}
\end{equation}
Using the dimensionless parameters defined in the
eqns.\eqref{denpara} and \eqref{dimexp} we get from the above
equation,
\begin{equation}\label{mod333}
E(z)=\sqrt{\frac{\left[\lambda\Omega_{m0}+\Omega_{rad0}\left(\lambda-m\right)\left(1+z\right)\right]\left(1+z\right)^{3}}{f_{0}\left(\lambda-m\right)\left(3H_{0}^{2}\lambda\Omega_{m0}\left(1+z\right)^{3}\right)^{m}}}
\end{equation}
In this model the working parameter space which is to be
constrained is $\left(\lambda, f_{0}, m\right)$.

\subsubsection{Model IV: Non-minimal coupling}
The model is given by,
\begin{equation}\label{mod4}
f(R,T)=R^{n}+f_{0}R^{n}T^{m}
\end{equation}
where $f_{0}\neq 0$, $n$ and $m$ are constant parameters. For this
model using the FLRW equations we get for $n=1$,

$$H^{2}(z)=\frac{\left[\lambda\left(\rho_{m0}+\rho_{rad0}\left(1+z\right)\right)+f_{0}\left(\lambda\rho_{m0}\left(1+z\right)^{3}\right)^{m}\left(\lambda\rho_{m0}+\rho_{rad0}\left(\lambda-m\right)\left(1+z\right)\right)\right]\left(1+z\right)^{3}}{\lambda+f_{0}\left(3m+\lambda\right)\left(\lambda\rho_{m0}\left(1+z\right)^{3}\right)^{m}} \times$$

\begin{equation}\label{mod44}
\left[\frac{4\lambda}{3\left\{\lambda+f_{0}\left(\lambda-m\right)\left(\lambda\rho_{m0}\left(1+z\right)^{3}\right)^{m}\right\}}-\frac{1}{1+f_{0}\left(\lambda\rho_{m0}\left(1+z\right)^{3}\right)^{m}}\right]
\end{equation}

Using the dimensionless parameters defined in the
eqns.\eqref{denpara} and \eqref{dimexp} we get from the above
equation,

$$E(z)=\frac{\left[3\lambda\left(\Omega_{m0}+\Omega_{rad0}\left(1+z\right)\right)+3f_{0}\left(H_{0}^{2}\right)^{m}\left(3\lambda\Omega_{m0}\left(1+z\right)^{3}\right)^{m}\left(\lambda\Omega_{m0}+\Omega_{rad0}\left(\lambda-m\right)\left(1+z\right)\right)\right]^{1/2}\left(1+z\right)^{3/2}}{\left[\lambda+f_{0}\left(3m+\lambda\right)\left(3H_{0}^{2}\lambda\Omega_{m0}\left(1+z\right)^{3}\right)^{m}\right]^{1/2}} \times$$

\begin{equation}\label{mod444}
\left[\frac{4\lambda}{3\left\{\lambda+f_{0}\left(\lambda-m\right)\left(3H_{0}^{2}\lambda\Omega_{m0}\left(1+z\right)^{3}\right)^{m}\right\}}-\frac{1}{1+f_{0}\left(3H_{0}^{2}\lambda\Omega_{m0}\left(1+z\right)^{3}\right)^{m}}\right]^{1/2}
\end{equation}
In this model the parameter space which is to be constrained is
$\left(\lambda, f_{0}, m\right)$.

\subsubsection{Model V: $\sqrt{-T}$ model (Shabani \& Farhoudi 2013)}
The model is given by,
\begin{equation}\label{mod5}
f(R,T)=\alpha R^{-n}+\sqrt{-T}
\end{equation}
where $\alpha>0$ and $n\neq 0$ are constant parameters. For this
model using the first FLRW equation we get for $n=-1$,
\begin{equation}\label{mod55}
H^{2}(z)=\frac{1}{6\alpha\lambda}\left[2\lambda\left(\rho_{m0}+\rho_{rad0}\left(1+z\right)\right)\left(1+z\right)^{3}-\left(\lambda-1\right)\sqrt{-\lambda\rho_{m0}\left(1+z\right)^{3}}\right]
\end{equation}
Using the dimensionless parameters defined in the
eqns.\eqref{denpara} and \eqref{dimexp} we get from the above
equation,
\begin{equation}\label{mod555}
E(z)=\frac{1}{\sqrt{2\alpha\lambda}}\left[2\lambda\left(\Omega_{m0}+\Omega_{rad0}\left(1+z\right)\right)\left(1+z\right)^{3}-\left(\lambda-1\right)\sqrt{-\left(3H_{0}^{2}\right)^{-1}\lambda\Omega_{m0}\left(1+z\right)^{3}}\right]^{1/2}
\end{equation}
Here the parameter space which is to be constrained is
$\left(\lambda, \alpha\right)$.

\subsection{The Data}
Here we will use the $30$ point $z-H(z)$ cosmic chronometer data
sets (Jimenez \& Loeb 2002; Moresco 2015; Simon, Verde \& Jimenez
2005; Stern et al. 2010; Zhang et al. 2014). The complete set of
the CC data has been presented in table \ref{table:7}. Cosmic
chronometers are very crucial parameters used to understand the
evolution history of the universe. We know that in an expanding
universe the most crucial factor is the expansion rate which is
represented by the Hubble parameter $H$. Cosmic chronometer is a
method that records the Hubble parameter data from the
observations of the early passively evolving galaxies. It uses the
technique of differential age evolution while retrieving the
Hubble data. It is known that we can represent the Hubble
parameter in terms of the redshift parameter $z$ as
$H=-\left(1+z\right)^{-1}dz/dt$. So mathematically speaking, we
may directly retrieve the Hubble parameter data by measuring the
time rate of change of the redshift parameter at a particular
redshift value. This technique was first introduced by Jimenez \&
Loeb 2002. After this the method became a popular means to perform
observational data analysis of theoretical models. The reader may
refer to various works related to cosmic chronometers in the refs.
Moresco (2015); Simon, Verde \& Jimenez (2005); Stern et al.
(2010); Zhang et al. (2014). The 30 point $z-H(z)$ CC data-set is
obtained in the redshift range of $0<z<2$ which spans over a
cosmic time of $10~ Gyr$. Moreover this data is captured in a
model independent way which makes it extremely suitable for
constraining the parameter spaces of theoretical models.

\subsection{Analysis with Cosmic Chronometer (CC) data}
In this section we would like to perform a data analysis with the
30 point cosmic chronometer data-set and constrain the parameter
space. The complete data-set is given in table \ref{table:7}. For
this purpose we will first establish the $\chi^{2}$ statistic as a
sum of standard normal distribution as follows:
\begin{equation}\label{chicc}
{\chi}_{CC}^{2}=\sum\frac{\left[H(z)-H_{obs}(z)\right]^{2}}{\sigma^{2}(z)}
\end{equation}
where $H(z)$ and $H_{obs}(z)$ are the theoretical and
observational values of Hubble parameter at different red-shifts
respectively and $\sigma(z)$ is the corresponding error in
measurement of the data point. The present value of Hubble
parameter is considered as $H_{0}$ = 72 $\pm$ 8 Km s$^{-1}$
Mpc$^{-1}$ and we also consider a fixed prior distribution for it.
The reduced chi square can be written as
\begin{equation}\label{reducechicc}
L=\chi_{R}^{2}= \int e^{-\frac{1}{2}{\chi}_{CC}^{2}}P(H_{0})dH_{0}
\end{equation}
where $P(H_{0})$ is the prior distribution function for $H_{0}$.

\subsection{Joint analysis with CC+BAO}
The BAO peak parameter may be defined by (Thakur, Ghose \& Paul
2009; Paul, Thakur \& Ghose 2010; Paul, Ghose \& Thakur 2011;
Ghose, Thakur \& Paul 2012):
\begin{equation}\label{baopeak}
{\cal A}=\frac{\sqrt{\Omega_{m}}}{E(z_{1})^{1/3}}
\left(\frac{1}{z_{1}}~\int_{0}^{z_{1}}
\frac{dz}{E(z)}\right)^{2/3}
\end{equation}
In the above expression $E(z)=H(z)/H_{0}$ is called the normalized
Hubble parameter. It is known from the SDSS survey that the
red-shift  $z_{1}=0.35$ is the prototypical value of red-shift
which we will consider in our analysis.

Now the $\chi^{2}$ function for the BAO measurement can be written
as

\begin{equation}\label{Eqn1.5.174}
\chi^{2}_{BAO}=\frac{({\cal A}-0.469)^{2}}{(0.017)^{2}}
\end{equation}
The total joint data analysis CC+BAO for the $\chi^{2}$ function
may be defined by (Thakur, Ghose \& Paul 2009; Paul, Thakur \&
Ghose 2010; Paul, Ghose \& Thakur 2011; Ghose, Thakur \& Paul
2012; Wu \& Yu 2007)
\begin{equation}\label{Eqn1.5.175}
\chi^{2}_{total}=\chi^{2}_{CC}+\chi^{2}_{BAO}
\end{equation}
Using the minimizing technique of the above $\chi^{2}_{total}$ we
would like to constrain the parameter space of the models. Here
the constraints are expected to be different from the constraints
obtained from the CC data because of the effects of the BAO peak
parameter in the system.

\subsection{Joint analysis with CC+BAO+CMB}
The first peak of the CMB power spectrum is basically a shift
parameter given by
\begin{equation}\label{Eqn1.5.176}
{\cal R}=\sqrt{\Omega_{m}} \int_{0}^{z_{2}} \frac{dz}{E(z)}
\end{equation}
where $z_{2}$ is the value of redshift corresponding to the last
scattering surface. From the WMAP 7-year data available in the
work of Komatsu et al. (2011) the value of the parameter has been
obtained as ${\cal R}=1.726\pm 0.018$ at the redshift $z=1091.3$.
Now the $\chi^{2}$ function for the CMB measurement can be written
as
\begin{equation}\label{Eqn1.5.177}
\chi^{2}_{CMB}=\frac{({\cal R}-1.726)^{2}}{(0.018)^{2}}
\end{equation}
Here we will consider the three cosmological tests together and
perform the joint data analysis for CC+BAO+CMB. The total
$\chi^{2}$ function for this case may be defined by
\begin{equation}\label{Eqn1.5.178}
\chi^{2}_{TOTAL}=\chi^{2}_{CC}+\chi^{2}_{BAO}+\chi^{2}_{CMB}
\end{equation}
Here in the presence of the CC data and the two peak parameters
(BAO and CMB) the parameter space of the models is supposed to be
tightly constrained which we will see in the next subsection.

\subsection{Constraints on the models from fitting analysis}
Now we will report the results that we have obtained from the
observational data analysis of the models with the different
data-sets. Here we have reported the best-fit values of the free
parameters along with the minimized $\chi^{2}$ value for all the
data-sets in the tabular form.As mentioned earlier, we have used
the publicly available \textit{CosmoMC} (Lewis et al. 2000; Lewis
\& Bridle 2002) for visualization and subsequent analysis. Using
\textit{CosmoMC}, we calculated the bounds of the free parameters
in $68\%$, $95\%$ and $99\%$ confidence intervals and generated 1D
distributions as well as 2D joint likelihood contours for the free
parameters.The distributions and the likelihood contours are
generated for all the three data-sets and depicted by different
colours in the figures. The different confidence levels are
represented by different shades of the same colour.

\subsubsection{ Constraints on Model-I}
For this model we have taken $m = 1$ and performed the fitting
analysis. The results are reported below in Tables \ref{table:1A}
and \ref{table:1B}.

\begin{table}[h!]
\centering
\begin{tabular} {l@{\hskip .3in}c@{\hskip .5in}c@{\hskip .5in} c@{\hskip .5in}  r}
\hline
Data   &  $\lambda$ & $\alpha$ & $\beta$ & ${\chi^2_{min}}$ \\
\hline
CC & $ -2.8571~ or ~-2.2857$ & $2.7143$ & $-0.3469~ or~ -0.3061$ & $6.1$ \\
CC+BAO & $-3.3061$ & $2.1428$& $-0.4285$ & $47.6055$ \\
CC+BAO+CMB & $-2.0816$ & $1.8163$ & $0.8367$ & $6783.9509$ \\
\hline
\end{tabular}
\vspace{3mm} \caption{The best fit values of $\lambda$, $\alpha$
and $\beta$, when $m = 1$ for Model-I, with the minimum values of
$\chi^2$} \label{table:1A}
\end{table}

\begin{table}[h!]
\centering
\begin{tabular} {l c@{\hskip .3in} c@{\hskip .3in} c@{\hskip .3in} r}
\hline
Parameter &~~ 68\% limits & 95\% limits & 99\% limits\\
\hline\\
$\lambda$  & $-2.93^{+0.87}_{-0.43}$  & $-2.93^{+0.91}_{-1.0}$ &  $-2.93^{+0.99}_{-1.1}$   \\
$\alpha$ & $1.82^{+0.41}_{-0.79}$  & $1.82^{+1.1}_{-0.87}$ & $1.82^{+1.2}_{-0.98}$  \\
$\beta$  & $-0.07^{+0.16}_{-0.26}$  & $-0.07^{+0.41}_{-0.34}$ &  $-0.07^{+0.53}_{-0.40}$
\\\\
\hline\\
$\lambda$  & $-3.00^{+0.59}_{-0.59}$  & $-3.00^{+0.96}_{-0.96}$ &  $-3.00^{+1.0}_{-1.0}$   \\
$H$ & $72.40^{+0.58}_{-0.58}$  & $72.40^{+0.96}_{-0.96}$ & $72.40^{+1.0}_{-1.0}$  \\
$\Omega_{m0}$  & $0.30^{+0.12}_{-0.12}$  & $0.30^{+0.19}_{-0.19}$ &  $0.30^{+0.20}_{-0.20}$
\\\\
\hline
\end{tabular}
\vspace{3mm} \caption{Bounds on the free parameters from CC data
for different confidence limits} \label{table:1B}
\end{table}

\begin{figure}[hbt!]
 \centering
\includegraphics[width=80mm]{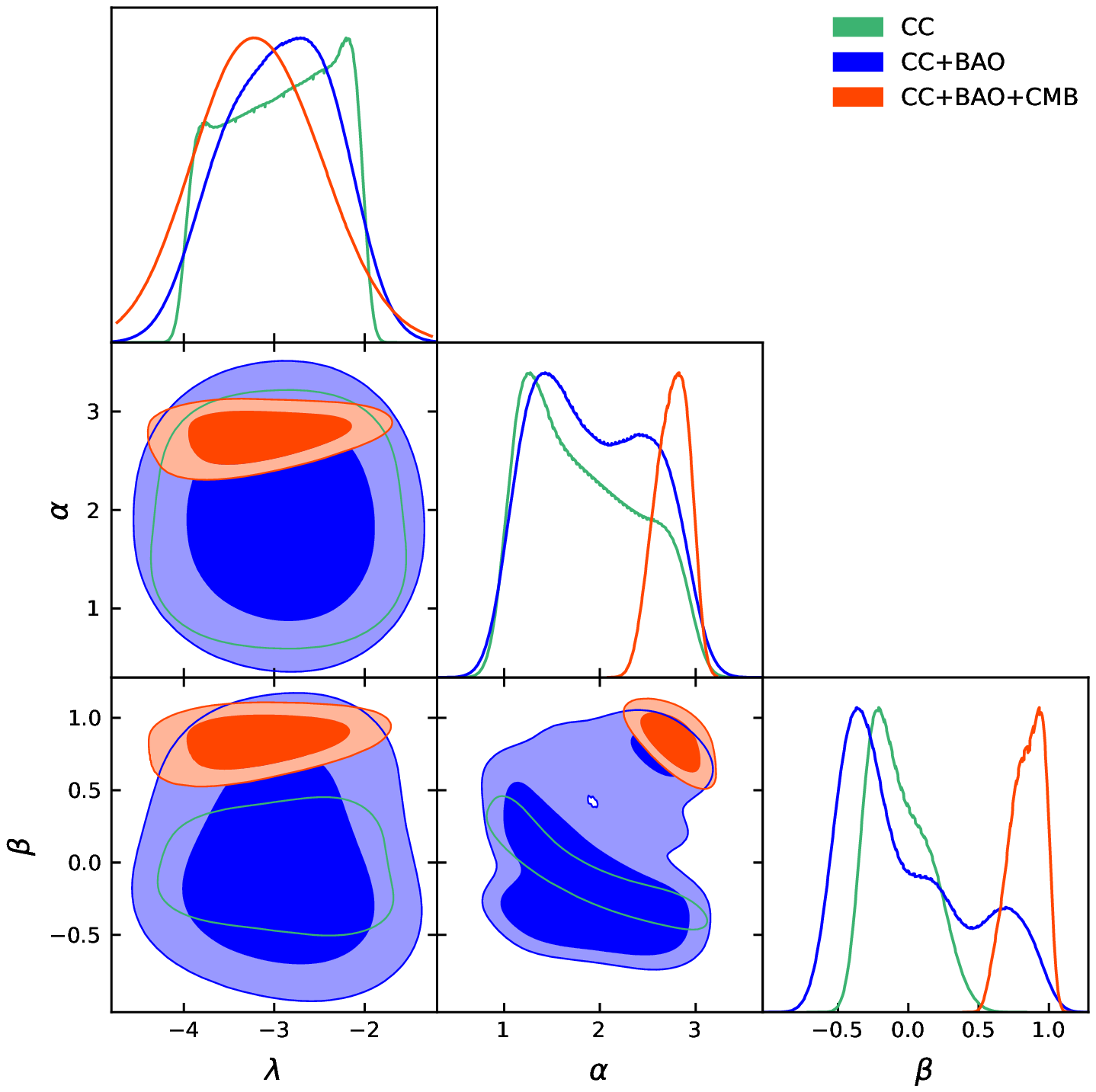}~~~~~~\includegraphics[width=80mm]{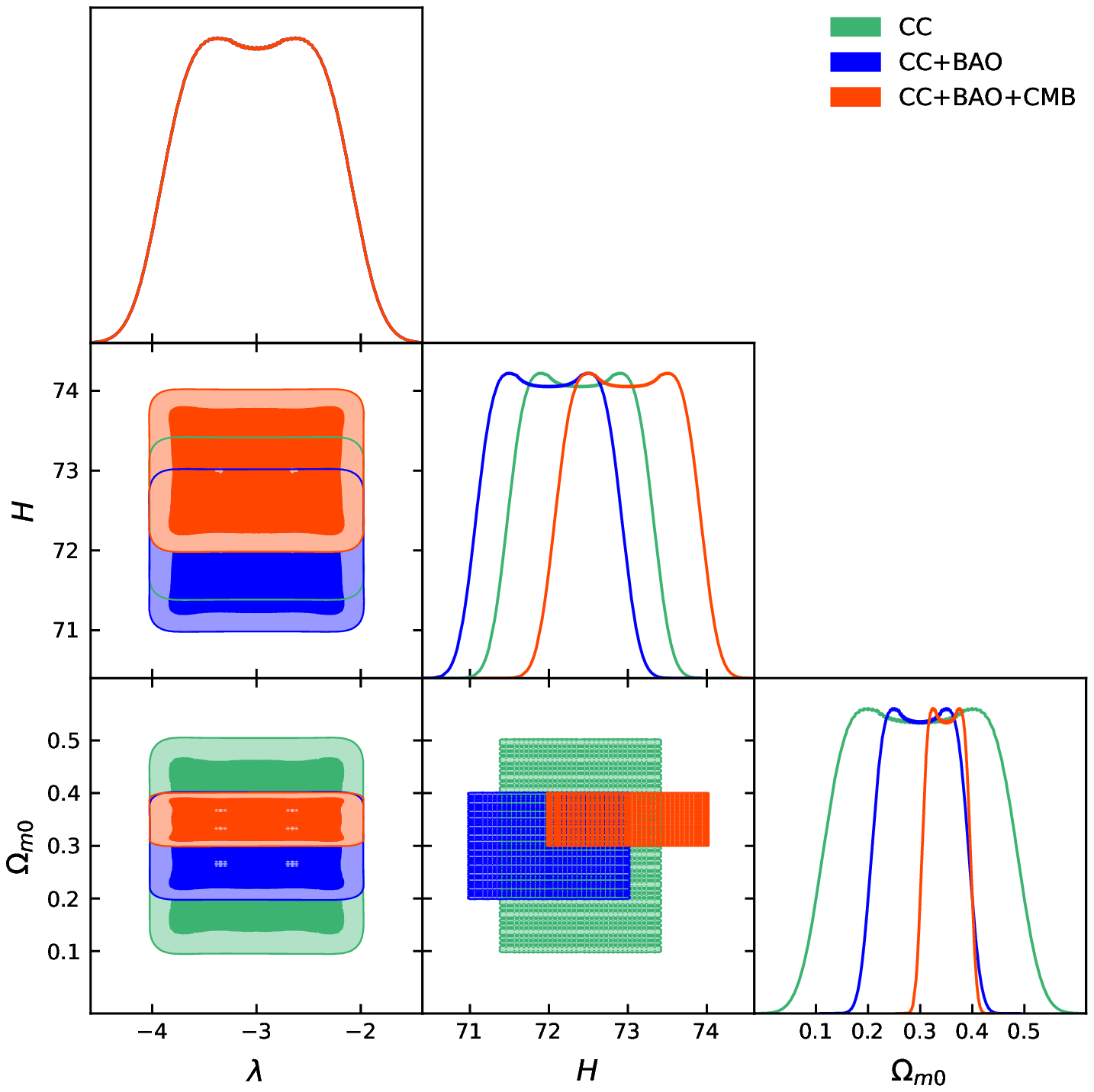}~~~~\\
\caption{1D distributions and 2D joint likelihood contours of the
free parameters ($\lambda, \alpha, \beta$) of Model I. The deeper
shades show the $68\%$ confidence intervals and the lighter shades
represent the $95\%$ confidence intervals for the parameters. The
figure on the right panel shows the likelihood contours for the
present day matter density parameter $\Omega_{m0}$ and present day
value of Hubble parameter $H$.}\label{f1}
\end{figure}
From Fig.\ref{f1}, it is evident that the distributions of the
free parameters are quite skewed compared to Gaussian
distribution. For the free parameter $\lambda$, we clearly see
that it's distribution is perfectly Gaussian centred around $-3$
for CC+BAO+CMB dataset. But for the other datasets the centre
shifts to the right and moreover the Gaussian nature is distorted.
In case of $\alpha$ we see that for CC+BAO+CMB data-set the
distribution is Gaussian around $2.8$. For the other data-sets the
distribution is distorted and the centre shifts towards the left.
The distribution of $\beta$ is not perfectly Gaussian for any
date-set. But for CC+BAO+CMB it is nearly Gaussian around 0.8. The
centre shifts towards left for the other data-sets. The different
confidence levels of the free parameters are shown in the contours
using different shades. From the contours we see that the
parameter limits obtained for CC+BAO data-sets are the least
constrained and those for CC+BAO+CMB are most constrained. The
constrained values for the parameters $\Omega_{m0}$ and $H$
presented in the table \ref{table:1B} are obtained in the
acceptable range according to the recent cosmological
observations.

\subsubsection{Constraints on Model-II}
In this model we have fixed $m = 1$ and performed the fitting
analysis. The results are summarized in the following tables,
viz., \ref{table:2A} and \ref{table:2B}.
\begin{table}[h!]
\centering

\begin{tabular} {l@{\hskip .3in}c@{\hskip .5in}c@{\hskip .5in} c@{\hskip .5in}  r}
\hline
Data   &  $\lambda$ & $\alpha$ & $\beta$ & ${\chi^{^2}_{min}}$ \\
\hline
CC & $-0.0082$ & $0.6836$ & $49.7959$ & $7.1$ \\
CC+BAO & $ -0.0041$ & $ 0.8469$& $-0.4285$ & $55.9183$ \\
CC+BAO+CMB & $-9.9e-05$ & $0.5102$ & $53.0612$ & $6640.0877$ \\
\hline
\end{tabular}
\vspace{3mm} \caption{The best fit values of $\lambda$, $\alpha$
and $\beta$, when $m = 1$ for Model-II, with the minimum values of
$\chi^2$} \label{table:2A}
\end{table}

\begin{table}[h!]
\centering
\begin{tabular} {l c@{\hskip .3in} c@{\hskip .3in} c@{\hskip .3in} r}
\hline
Parameter &~ 68\% limits & 95\% limits & 99\% limits\\
\hline\\
$\lambda$  & $-0.102^{+0.058}_{-0.0.058}$  & $-0.102^{+0.094}_{-0.094}$ &  $-0.102^{+0.098}_{-0.098}$   \\
$\alpha$ & $0.74^{+0.11}_{-0.13}$  & $0.74^{+0.23}_{-0.21}$ & $ 0.74^{+0.26}_{-0.24}$  \\
$\beta$  & $49.97^{+5.98}_{-5.98}$  & $49.99^{+9.88}_{-9.88}$ &  $49.99^{+9.91}_{-9.91}$   \\
\hline\\\\
$\lambda$  & $-1.00^{+0.59}_{-0.59}$  & $-1.00^{+0.96}_{-0.96}$ &  $-1.00^{+1.0}_{-1.0}$   \\
$H$ & $72.60^{+0.35}_{-0.35}$  & $72.60^{+0.58}_{-0.58}$ & $72.60^{+0.60}_{-0.60}$  \\
$\Omega_{m0}$  & $0.35^{+0.029}_{-0.029}$  &
$0.35^{+0.048}_{-0.048}$ & $0.35^{+0.050}_{-0.050}$
\\\\
\hline
\end{tabular}
\vspace{3mm} \caption{Bounds on the free parameters from CC data
for different confidence limits} \label{table:2B}
\end{table}

\begin{figure}[hbt!]
 \centering
 \includegraphics[width=80mm]{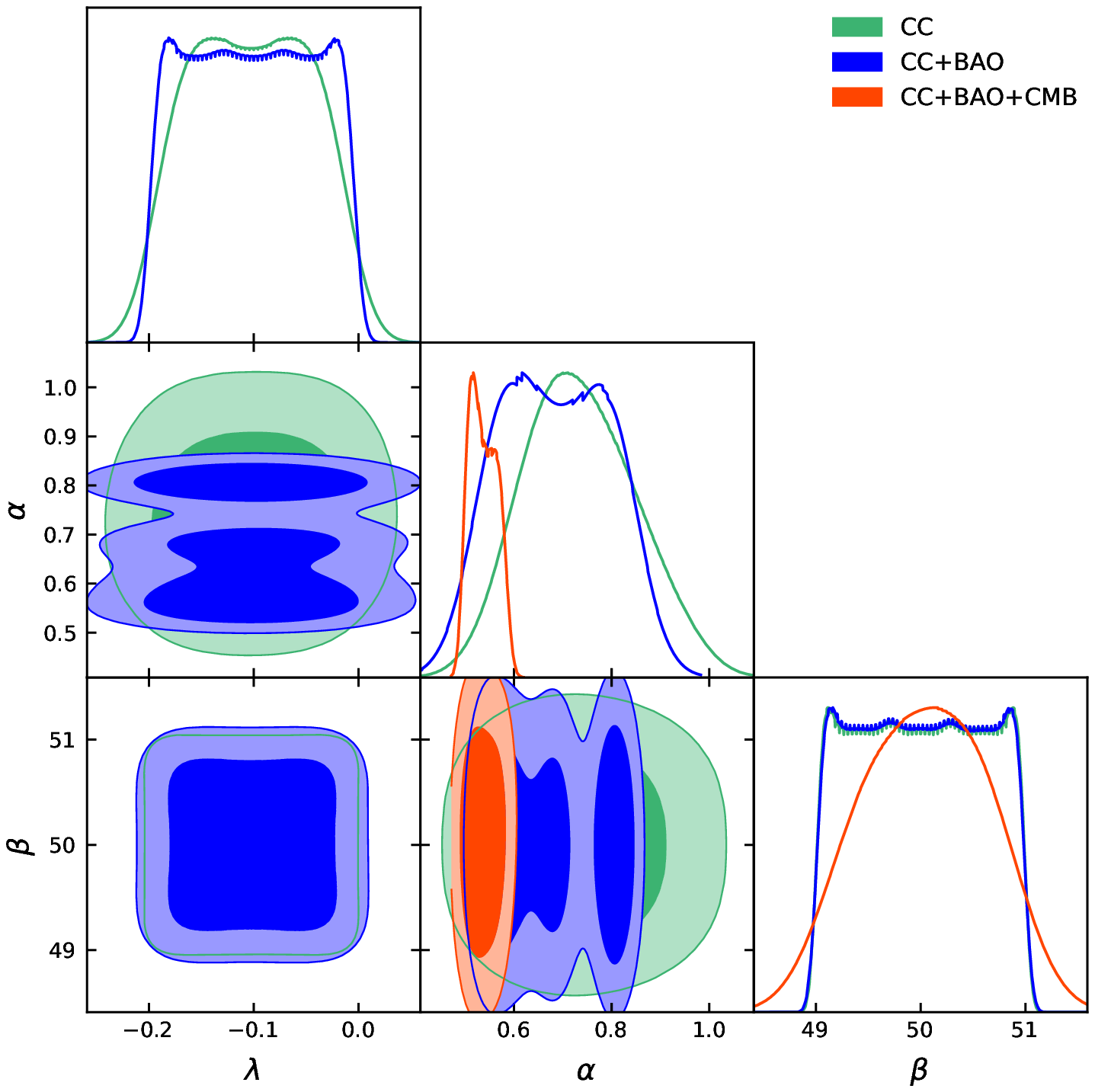}~~~~~~\includegraphics[width=80mm]{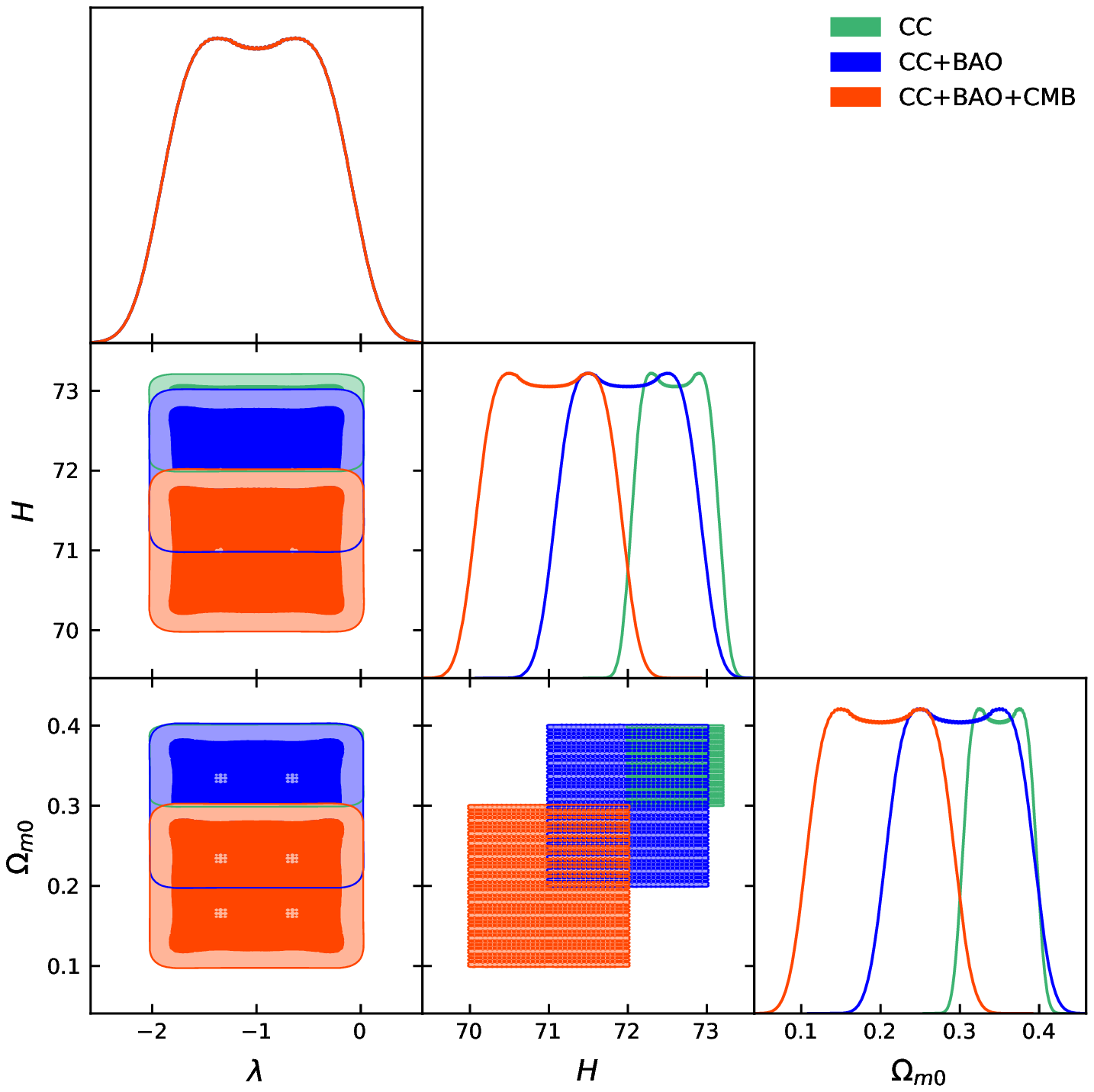}~~~~\\
\caption{1D distributions and 2D joint likelihood contours of the
free parameters ($\lambda, \alpha, \beta$) of Model II. The deeper
shades show the $68\%$ confidence intervals and the lighter shades
represent the $95\%$ confidence intervals for the parameters. The
figure on the right panel shows the likelihood contours for the
present day matter density parameter $\Omega_{m0}$ and present day
value of Hubble parameter $H$.}\label{f2}
\end{figure}
Figure \ref{f2} represents the confidence contours for this model.
From the distribution plots of the free parameters we see that the
distributions are highly skewed compared to the Gaussian
distribution. For $\lambda$ there is no proper peak of the
distribution and there is no well defined distribution for the
CC+BAO+CMB case. For $\alpha$, the distributions are comparatively
smoother with a proper Gaussian centred at $0.74$ for the CC data
set. For the other data-sets there are significant distortions
evident from the plots. Finally for $\beta$ we see that the
distributions for CC and CC+BAO are almost identical. The
distribution for CC+BAO+CMB is comparatively skewed towards the
right. Moreover the contours involving $\lambda$ does not show the
case for CC+BAO+CMB. This is probably due to computational
complicacy of the model. Due to the involvement of the exponential
function probably there is no real limit for $\lambda$ for this
data-set. Unlike the previous model here we see that the contours
for CC data-set are least constrained and those for CC+BAO+CMB
(for $\alpha$ only) are the most constrained scenarios.

\subsubsection{Constraints on Model-III}
Now, we will turn our focus on model-III. Unlike Model-I and
Model-II, here, we did not require to fix any free parameters
since the number of free parameters are quite less and manageable.
In the tables \ref{table:3A} and \ref{table:3B}, the details of
the results are summarized.
\begin{table}[h!]
\centering
\begin{tabular} {l@{\hskip .3in}c@{\hskip .5in}c@{\hskip .5in} c@{\hskip .5in}  r}
\hline
Data   &  $\lambda$ & $f0$ & $m$ & ${\chi^{2}_{min}}$ \\
\hline
CC & $0.3332$ & $-1.8775$ & $0.3387$ & $5.4$\\
CC+BAO & $ 0.6969$ & $-1.1836$& $0.7244$ & $33.1537$ \\
CC+BAO+CMB & $0.1716$ & $-1.6122$ & $0.1734$ & $6191.3833$ \\
\hline
\end{tabular}
\vspace{3mm} \caption{The best fit values of $\lambda$, $f0$ and
$m$ for Model-III, with the minimum values of $\chi^2$}
\label{table:3A}
\end{table}

\begin{table}[h!]
\centering
\begin{tabular} {l c@{\hskip .3in} c@{\hskip .3in} c@{\hskip .3in} r}
\hline
Parameter &~~ 68\% limits & 95\% limits & 99\% limits\\
\hline\\
$\lambda$   & $0.273^{+0.070}_{-0.11}$  & $0.27^{+0.30}_{-0.18}$ &  $0.27^{+0.34}_{-0.22}$   \\
$f0$ &  $ -1.42^{+0.38}_{-0.25}$  & $-1.42^{+0.48}_{-0.56}$ & $  -1.42^{+0.56}_{-0.66}$  \\
$m$  & $0.286^{+0.062}_{-0.098}$  & $0.29^{+0.31}_{-0.19}$ &  $0.29^{+0.32}_{-0.21}$
\\\\
\hline
\\
$\lambda$  & $0.25^{+0.15}_{-0.15}$  & $0.25^{+0.24}_{-0.24}$ &  $0.25^{+0.25}_{-0.25}$   \\
$H$ & $73.0^{+0.58}_{-0.58}$  & $73.0^{+0.96}_{-0.96}$ & $73.0^{+1.0}_{-1.0}$  \\
$\Omega_{m0}$  & $0.40^{+0.12}_{-0.12}$  & $0.40^{+0.19}_{-0.19}$
& $0.40^{+0.20}_{-0.20}$
\\\\
\hline
\end{tabular}
\vspace{3mm} \caption{Bounds on the free parameters from CC data
for different confidence limits} \label{table:3B}
\end{table}

\begin{figure}[hbt!]
 \centering
 \includegraphics[width=80mm]{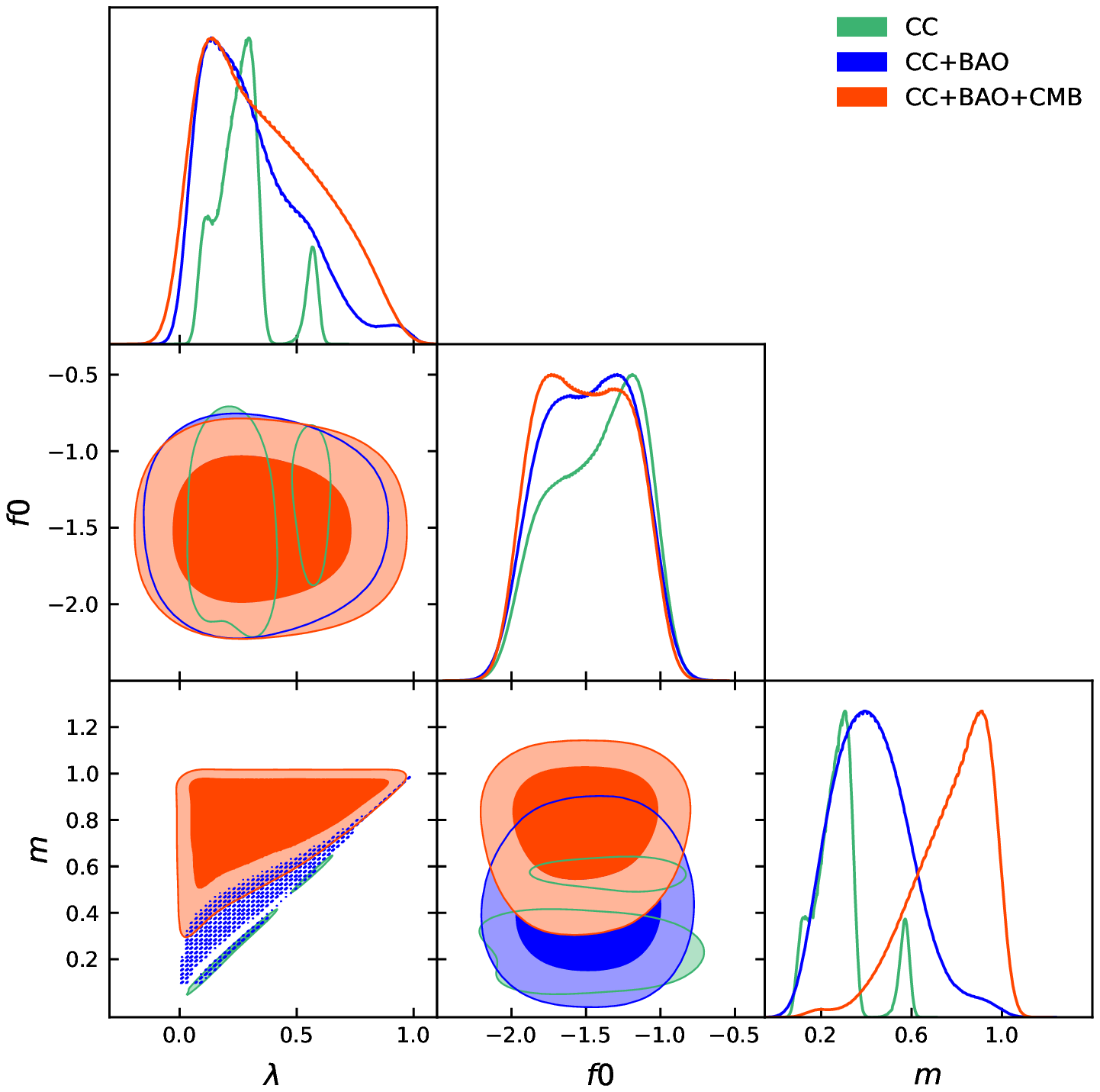}~~~~~~\includegraphics[width=80mm]{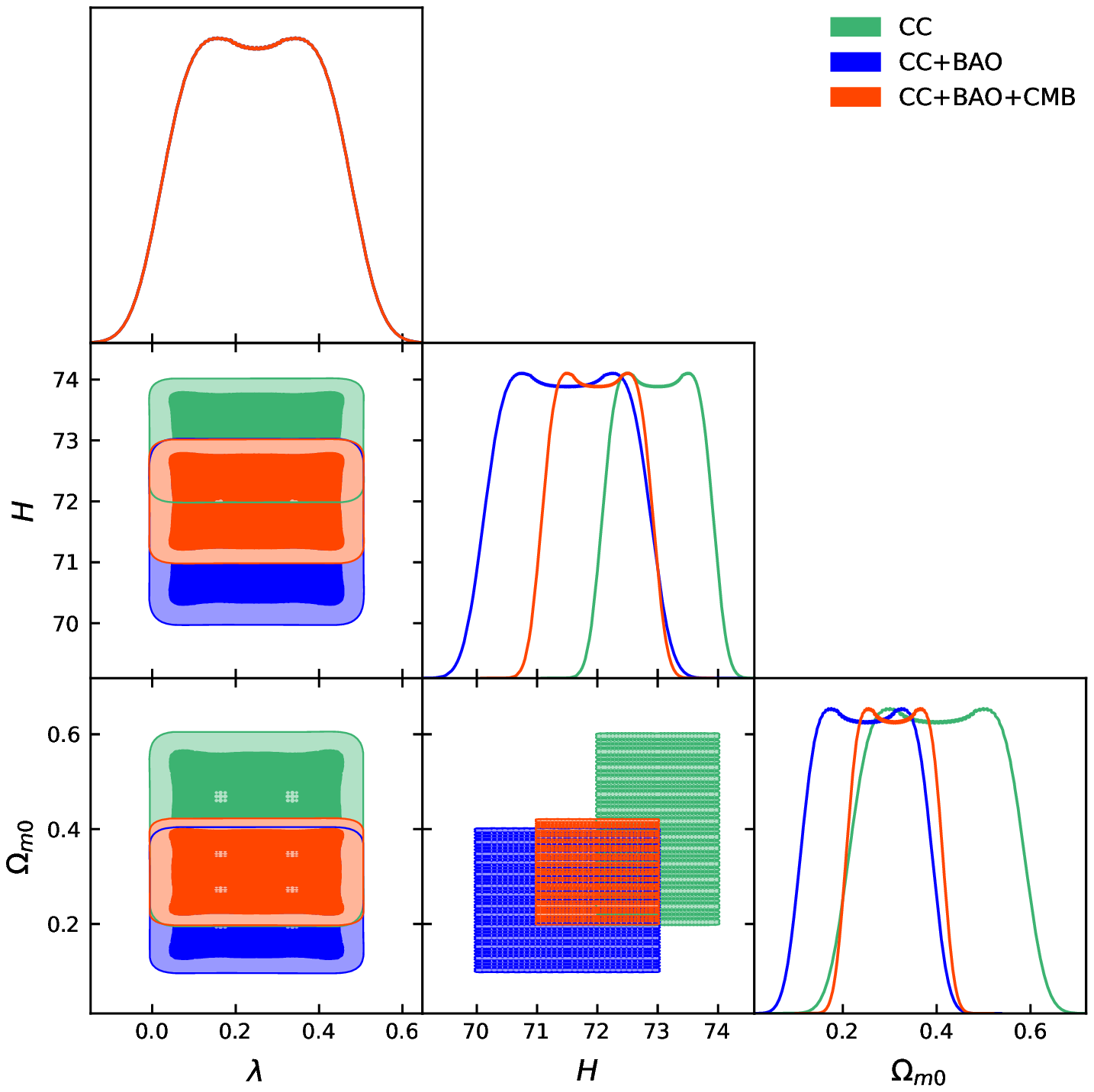}~~~~\\
\caption{1D distributions and 2D joint likelihood contours of the
free parameters ($\lambda, f_{0}, m$)  of Model III. The deeper
shades show the $68\%$ confidence intervals and the lighter shades
represent the $95\%$ confidence intervals for the parameters. The
figure on the right panel shows the likelihood contours for the
present day matter density parameter $\Omega_{m0}$ and present day
value of Hubble parameter $H$.}\label{f3}
\end{figure}
Figure \ref{f3} shows the 2D confidence contours and the
distributions followed by the free parameters of model-III. From
the distribution curves we see that for none of the parameters we
get a perfect Gaussian distribution for any data-set. The nearest
is the distribution for $m$ for the CC+BAO dataset, which is
almost Gaussian in nature with some irregularities in the right
wing. The other curves for $m$ are quite skewed on either sides.
Similarly the distributions for $\lambda$ are quite skewed
compared to a normal curve. The situation for $f_{0}$ is better
with bell shaped curves but does not possess a well defined peak.
The contours give the ranges of the parameters in 2D scenario for
$68\%$ and $95\%$ confidence limits. For this model the most
constrained parameter limits are obtained for the CC data-set,
which is contrary to the results obtained in the previous models.
The least constrained contours are obtained for the CC+BAO+CMB
data-set. For this model the constrained value of $\Omega_{m0}$
obtained, is on a little higher side compared to the
cosmologically accepted range.

\subsubsection{Constraints on Model-IV}
Here we will report the results obtained from the analysis for
model-IV. The model is fitted with the observational data via the
$\chi^2$ minimization mechanism and the best fit values of the
free parameters are estimated. Ranges of parameters and their
confidence contours for different limits are generated using the
\textit{CosmoMC} code. The results are reported in the following
tables,\ref{table:4A} and \ref{table:4B}.

\begin{table}[h!]
\centering
\begin{tabular} {l@{\hskip .3in}c@{\hskip .5in}c@{\hskip .5in} c@{\hskip .5in}  r}
\hline
Data   &  $\lambda$ & $f0$ & $m$ & ${\chi^{2}_{min}}$ \\
\hline
CC & $0.3265$ & $ 2.6122$ & $0.3327$ & $2.2$\\
CC+BAO & $0.1836$ & $2.4285$ & $0.1002$ & $310.1417$ \\
CC+BAO+CMB & $0.1632$ & $ 2.3877$ & $0.1775$ & $4920.1211$ \\
\hline
\end{tabular}
\vspace{3mm} \caption{The best fit values of $\lambda$, $f0$ and
$m$ for Model-IV, with the minimum values of $\chi^2$}
\label{table:4A}
\end{table}

\begin{table}[h!]
\centering
\begin{tabular} {l  c@{\hskip .3in} c@{\hskip .3in} c@{\hskip .3in} r}
\hline
Parameter  &~~ 68\% limits & 95\% limits & 99\% limits\\
\hline\\
$\lambda$  &  $0.258^{+0.11}_{-0.090}$  & $0.26^{+0.15}_{-0.17}$ &  $0.26^{+0.15}_{-0.17}$   \\
$f0$ & $2.43^{+0.29}_{-0.34}$  & $2.43^{+0.60}_{-0.56}$ & $2.43^{+0.75}_{-0.70}$  \\
$m$  & $0.266^{+0.11}_{-0.085}$  & $0.27^{+0.14}_{-0.17}$ &  $0.27^{+0.14}_{-0.17}$
\\\\
\hline
\\
$\lambda$  & $0.25^{+0.15}_{-0.15}$  & $0.25^{+0.24}_{-0.24}$ &  $0.25^{+0.25}_{-0.25}$   \\
$H$ & $72.50^{+0.87}_{-0.87}$  & $72.50^{+1.4}_{-1.4}$ & $72.50^{+1.5}_{-1.5}$  \\
$\Omega_{m0}$  & $0.27^{+0.16}_{-0.16}$  & $0.27^{+0.22}_{-0.22}$
& $0.27^{+0.24}_{-0.24}$
\\\\
\hline
\end{tabular}
\vspace{3mm} \caption{Bounds on the free parameters from CC data
for different confidence limits} \label{table:4B}
\end{table}

\begin{figure}[hbt!]
\centering
\includegraphics[width=80mm]{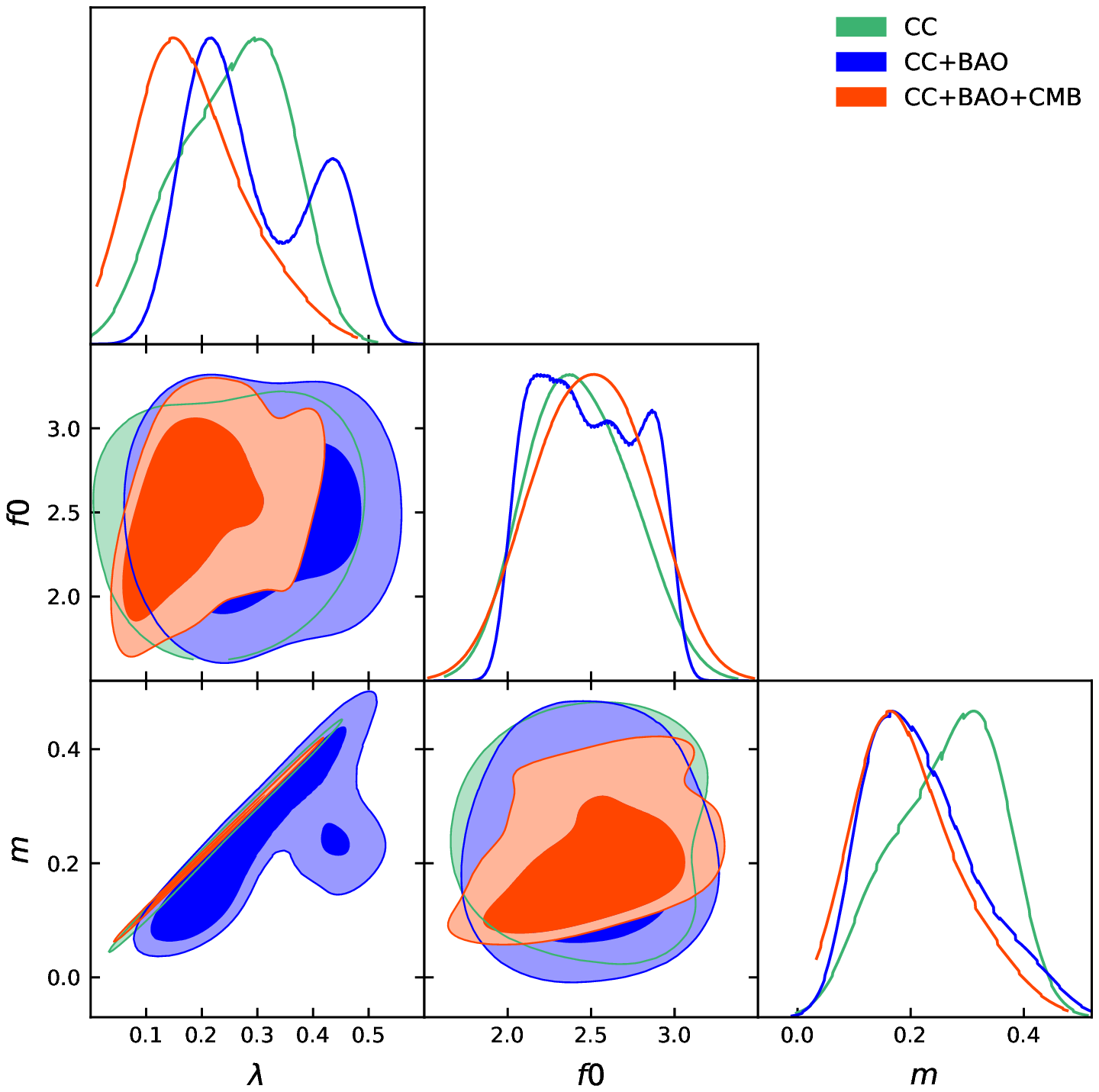}~~~~~~\includegraphics[width=80mm]{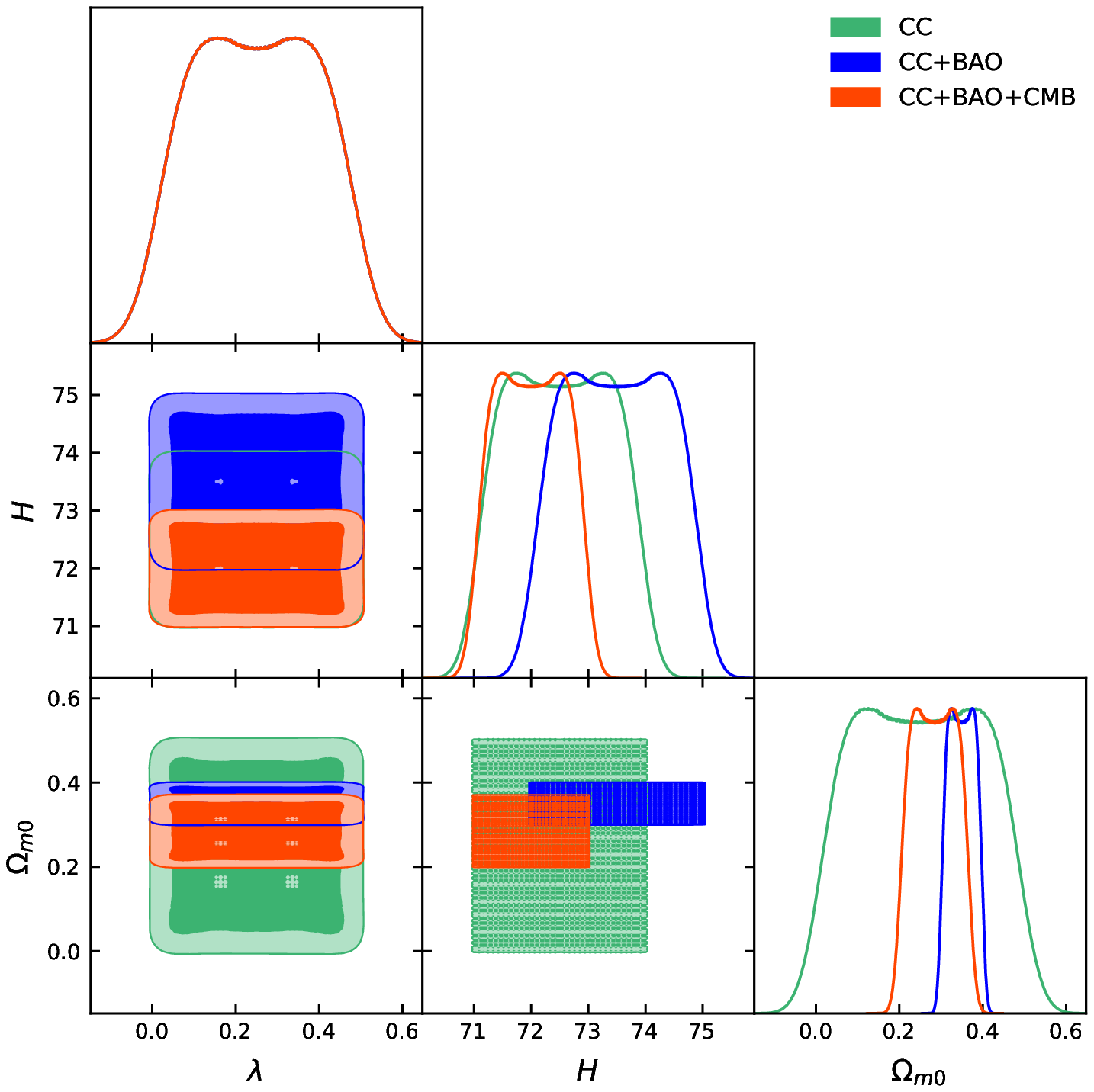}~~~~\\
\caption{1D distributions and 2D joint likelihood contours of the
free parameters ($\lambda, f_{0}, m$)  of Model IV. The deeper
shades show the $68\%$ confidence intervals and the lighter shades
represent the $95\%$ confidence intervals for the parameters. The
figure on the right panel shows the likelihood contours for the
present day matter density parameter $\Omega_{m0}$ and present day
value of Hubble parameter $H$.}\label{f4}
\end{figure}
In fig.\ref{f4} we have generated the 2D confidence contours and
the distributions for the free parameters of model-IV. Here the
parameter distributions are far more Gaussian like than the
previous model. Specially for $f_{0}$ the distributions for CC and
CC+BAO+CMB data-sets are perfectly Gaussian with a little
difference between the means. For this model the distributions for
CC+BAO data-set show the greatest skewness compared to Normal
distribution at least for $\lambda$ and $f_{0}$. For $m$ the
scenarion is relatively smoother, but not completely devoid of
skewness. Here we see that the parameter space is tightly
constrained by the CC+BAO+CMB data-set. The other data-sets apply
comparatively lighter grips on the parameters.

\subsubsection{Constraints on Model-V}
Finally, we will report the result of our fifth model here. It is
relevant to mention here that this model has some background
theoretical motivations and hence the results for this model may
be interesting for referencing other results. Here, we have only
two free parameters, which is computationally a convenient
scenario. The results are given in the tables \ref{table:5A} and
\ref{table:5B}.
\begin{table}[h!]
\centering
\begin{tabular} {l@{\hskip .1in}c@{\hskip .5in} c@{\hskip .5in}  r}
\hline
Data   &  $\lambda$ & $\alpha$  & ${\chi^{2}_{min}}$ \\
\hline
CC & $-0.3279$ & $0.3941$  & $3.84$\\
CC+BAO & $-0.2014$ & $0.1119$ & $37.9462$ \\
CC+BAO+CMB & $-0.1862$ & $0.0766$ & $6447.6411$ \\
\hline
\end{tabular}
\vspace{3mm} \caption{The best fit values of $\lambda$ and
$\alpha$ for Model-V, with the minimum values of $\chi^2$}
\label{table:5A}
\end{table}

\begin{table}[h!]
\centering
\begin{tabular} {l  c@{\hskip .3in} c@{\hskip .3in} c@{\hskip .3in} r}
\hline
Parameter &~~ 68\% limits & 95\% limits & 99\% limits\\
\hline\\
$\lambda$  & $-0.25^{+0.14}_{-0.14}$  & $-0.25^{+0.24}_{-0.24}$ &  $-0.25^{+0.25}_{-0.25}$   \\

$\alpha$  &  $0.408^{+0.057}_{-0.045}$  & $0.408^{+0.087}_{-0.093}$ &  $0.408^{+0.092}_{-0.11}$
\\\\
\hline
\\
$\lambda$  & $-0.25^{+0.15}_{-0.15}$  & $-0.25^{+0.24}_{-0.24}$ &  $0.25^{+0.26}_{-0.26}$   \\
$H$ & $71.50^{+0.29}_{-0.29}$  & $71.50^{+0.48}_{-0.48}$ & $71.50^{+0.50}_{-0.50}$  \\
$\Omega_{m0}$  & $0.20^{+0.059}_{-0.059}$  &
$0.20^{+0.096}_{-0.096}$ & $0.20^{+0.10}_{-0.10}$
\\\\
\hline
\end{tabular}
\vspace{3mm} \caption{Bounds on the free parameters from CC data
for different confidence limits} \label{table:5B}
\end{table}

\begin{figure}[hbt!]
\centering
\includegraphics[width=70mm]{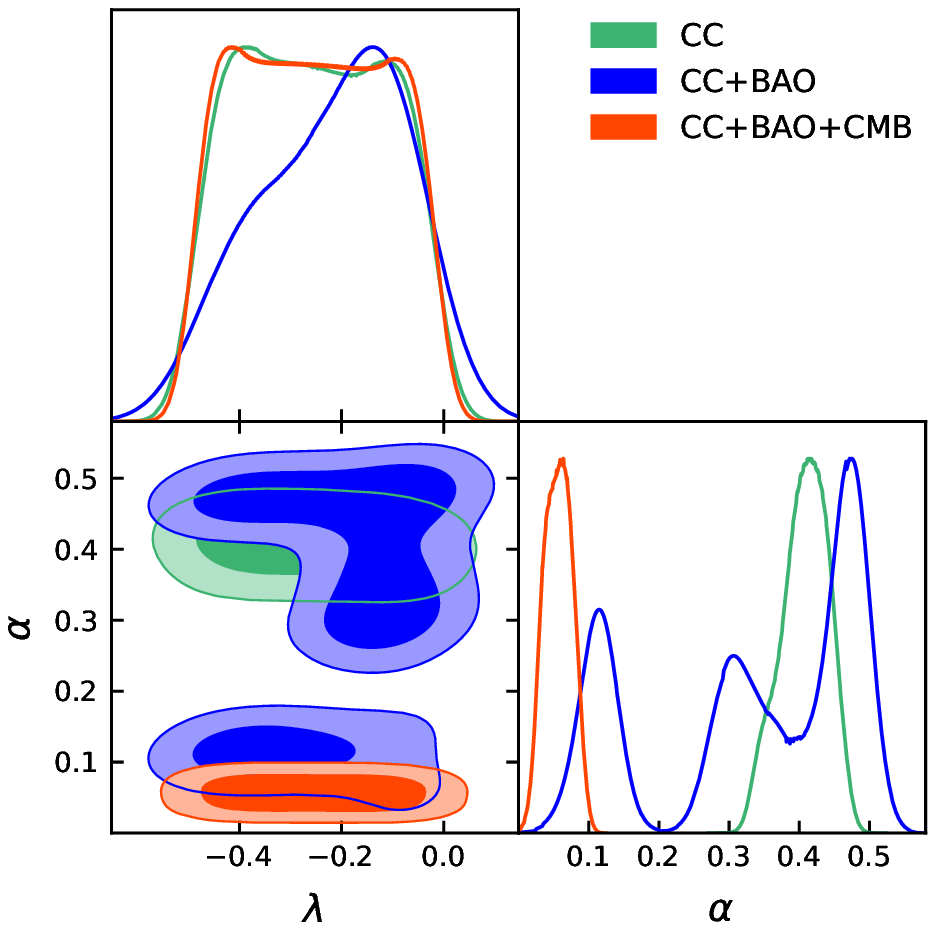}~~~~~~\includegraphics[width=80mm]{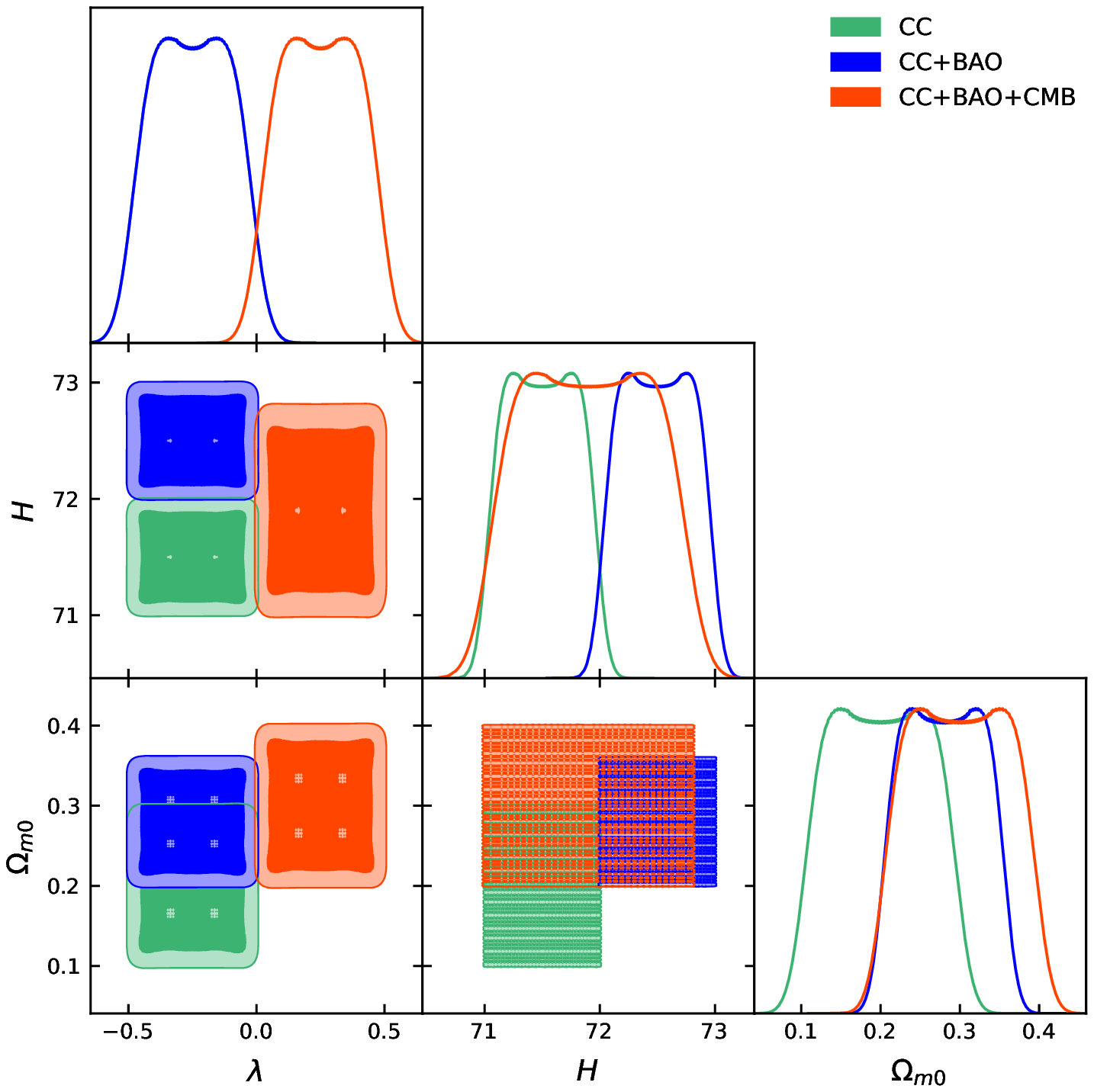}~~~~\\
\caption{1D distributions and 2D joint likelihood contours of the
free parameters ($\lambda, \alpha$) of Model V. The deeper shades
show the $68\%$ confidence intervals and the lighter shades
represent the $95\%$ confidence intervals for the parameters. The
figure on the right panel shows the likelihood contours for the
present day matter density parameter $\Omega_{m0}$ and present day
value of Hubble parameter $H$.}\label{f5}
\end{figure}

In fig.\ref{f5} we have shown the 2D confidence contours and the
distributions for the free parameters of model-V. We see that both
the parameters behave strangely for the CC+BAO data-set. This was
seen in the previous model as well. The distribution of $\alpha$
is highly scattered over a large interval for the CC+BAO data-set.
This shows that it is very lightly constrained by this data. The
results for the other data sets are better. This fact is clearly
seen in the contours where the intervals for the CC+BAO data sets
are highly de-localized. In fact there are two different ranges at
two different levels for the CC+BAO data-set. This is very unique
and not seen in any other models. The parameter space is tightly
constrained by the CC+BAO+CMB data-set. For this model the
constrained value of $\Omega_{m0}$ obtained, is on a little lower
side compared to the cosmologically accepted range.

\section{Model comparison}
Here we will use two model comparison criteria to check our
models. They are the Akaike Information Criterion (AIC) (Akaike
1974) and the Bayesian or Schwarz information criterion (BIC)
(Schwarz 1978). These are defined as follows,
\begin{equation}
AIC=-2\ln\mathcal{L}+2d=\chi^{2}_{min}+2d
\end{equation}
and
\begin{equation}
BIC=-2\ln\mathcal{L}+d\ln N=\chi^{2}_{min}+d\ln N
\end{equation}
where, $\mathcal{L}=\exp{\left(-\chi^{2}_{min}/2\right)}$ is the maximum likelihood function, $d$ is the number of model parameters and $N$ represents the total number of data points in the data used to constrain the model parameters. These model selection criteria involves the number of free parameters used in the model which is a very important fact. This is because we know that a model having more number of free parameters has greater degree of freedom to fit an observational data by changing its shape more conveniently. But from the Occam's Razor we also know that a model with the least number of free parameters is most desirable for solving any problem. So a model with more number of free parameters needs to be imposed with some penalties in order to build up a logical criteria. This idea has been taken into account while building up these statistical relations for model testing. Now since AIC and BIC may take on any positive values, we need to have a reference model with respect to which the selection criteria must be framed. Here we will use the $\Lambda CDM$ model as our reference model in this comparison. So for any model denoted by $M$ we define $\Delta AIC=AIC_{M}-AIC_{\Lambda CDM}$ and $\Delta BIC=BIC_{M}-BIC_{\Lambda CDM}$. Now using this difference parameter we have three cases according to the Jeffrey's scale.\\

i) When $\Delta AIC \leq 2$ or $\Delta BIC \leq 2$, then the model under scrutiny has good support from the reference model.\\

ii) When $4 \leq \Delta AIC \leq 7$ or $4 \leq \Delta BIC \leq 7$, then the model has lesser support from the reference model.\\

iii) When $\Delta AIC \geq 10$ or $\Delta BIC \geq 10$, then the model has no observational data support from the reference model.\\

Now it must be mentioned that these scales are not exclusive and
extreme care must be taken while using these (Nesseris \& Garcia
Bellido, 2013).

\begin{table}[h!]
\centering
\begin{tabular}{ |c|c|c|c|c|c|c| }
\hline
Models & Data-set & AIC &    $\Delta$AIC   & BIC   &   $\Delta$BIC \\
\hline \hline

\multirow{3}{4em}{$\Lambda CDM$ Model} & CC & 20.6068 & 0 & 48.216 & 0 \\
& CC+BAO & 55.1533 & 0  & 71.3601  & 0 \\
& CC+BAO+CMB & 577.50 & 0  &  645.93 & 0 \\
\hline

\multirow{3}{4em}{Model-I} & CC & 26.14 & 5.5332 & 73.6185 & 25.4025 \\
& CC+BAO & 65.3055 & 10.1522  & 95.6325  & 24.2724 \\
& CC+BAO+CMB & 650.75 & 73.25  &  693.21 & 47.28 \\
\hline

\multirow{3}{4em}{Model-II} & CC & 24.69 & 4.0832 & 75.3081 & 27.0921 \\
& CC+BAO & 69.9487 & 14.7954  & 97.3256  & 25.9655 \\
& CC+BAO+CMB & 654.39 & 76.89  & 689.92  & 43.99 \\
\hline

\multirow{3}{4em}{Model-III} & CC & 23.05 & 2.4432 & 55.1089 & 6.8929 \\
& CC+BAO & 60.1537 & 5.0004  & 82.1141  & 10.754 \\
& CC+BAO+CMB & 597.38 & 19.88  & 660.01  & 14.08 \\
\hline

\multirow{3}{4em}{Model-IV} & CC & 21.26 & 0.6532 & 59.9084 & 11.6924\\
& CC+BAO & 60.5156 & 5.3623  & 80.5621  & 9.202 \\
& CC+BAO+CMB & 591.72 & 14.22  & 658.73  & 12.80 \\
\hline

\multirow{3}{4em}{Model-V} & CC & 24.25 & 3.6432 & 57.3078 & 9.0918\\
& CC+BAO & 71.9472 & 16.7939  & 82.4562  & 11.0961 \\
& CC+BAO+CMB & 601.44 & 23.94  & 662.15  & 16.22 \\
\hline
\end{tabular}
\vspace{8mm}

\caption{The values for the AIC and BIC criteria for the different
models} \label{table:6}
\end{table}

The values of AIC and BIC criteria for all the models are reported
in the Table\ref{table:6}. From the table we see that for Model-I,
there is good observational support from CC data compared to
$\Lambda CDM$ model both from AIC and BIC criteria. But for CC+BAO
and CC+BAO+CMB setup there is almost no observational support both
according to AIC and BIC. Similarly for all the models there is
good support from CC data according to AIC and BIC criteria
compared to $\Lambda CDM$ model. Model-II has no support for
CC+BAO and CC+BAO+CMB just like Model-I. So Model-I and Model-II
are almost similar as far as support from observational data is
concerned. This can be attributed to the fact that both involve
coupling between matter and curvature in minimal form.

Model-III and IV both involve non-minimal coupling between $R$ and
$T$, but the support from observational data according to AIC and
BIC criteria are not analogous. We see that for Model-III there is
good support from CC+BAO dataset especially according to BIC
criteria. But for CC+BAO+CMB setup there is hardly any support.
For Model-IV the situation is just the reverse of model-III. Here
there is no support from CC+BAO dataset but there is very good
support from CC+BAO+CMB dataset for both the criteria. For Model-V
we see that there is observational support from CC+BAO according
to AIC criterion, but not according to BIC criterion. There is no
support from CC+BAO+CMB dataset for this model. So it is seen that
all the models are having observational support from some of the
data setups but none is having support from all the data types.
From this we can conclude that none of the models can be ruled out
from this analysis. All are having certain degree of efficiency as
far as complying with the observational data is concerned.
Moreover looking at the AIC and BIC values it can be concluded
that Model-III, IV and V are favoured over the models-I and II.
This again confirms the fact that non-minimal coupling is a
cosmological favoured set-up. However it should be kept in mind
that this is based on the fact that the comparison is done with
respect to the $\Lambda CDM$ model which we have considered as the
reference model. Taking a different reference or data-set may
change our conclusion. However, we believe that our references
($\Lambda CDM$ model and CC data) used here are quite efficient
and accepted structures of contemporary cosmology. In such cases,
our analysis, results and conclusion should be quite reliable.

\section{Conclusion}
In this work we have performed an observational data analysis on
$f(R,T)$ gravity using the cosmic chronometer data. Five different
models of the gravity theory were considered taking into account
both minimal and non-minimal coupling between matter and geometry.
The first two models were constructed based on minimal coupling
using the power law and exponential models. The third model
involved pure non-minimal coupling between matter and geometry and
the fourth model was the non-minimal coupling model. Finally we
picked up the fifth model from the literature which makes it the
most motivated model of all the models probed. The first four
models were constructed in such a way that they covered almost all
mathematical possibilities. This was achieved by keeping the
models generic in nature so that all other possibilities will be
some limiting or particular cases of these models. We have used
the 30 point $z-H(z)$ cosmic chronometer data to constrain the
free parameters of these models. We constructed the $\chi^{2}$
statistic using the Hubble parameter value from the data and the
theories. We have considered three different data settings namely,
CC, CC+BAO and CC+BAO+CMB using which the analysis is done. Adding
the BAO and CMB peak parameters with the CC data setting gave an
extra edge and helped us to get better constraints on the
parameters. Using a minimizing technique we fitted the models with
the data and obtained the best fit values of the free parameters
of the models. Lesser the minimum $\chi^{2}$ value better is the
compliance of the theory with the data. Using the publicly
available \textit{CosmoMC} code we determined the acceptable
ranges of the free parameters in the respective models in three
different confidence intervals, i.e., $68\%$, $95\%$ and $99\%$
corresponding to the observational data. We have also generated
the confidence contours showing these ranges for all the free
parameters. The plots also show the general distribution followed
by the model parameters. All these have been done for all the
three different data-sets as can be seen from the tables and the
contours. Values for all the cases have been reported in the paper
and corresponding contour plots for all the confidence intervals
have been shown. It was seen that different data-sets showed
different degree of parameter constraining for different models.
Looking at the general distribution followed by the parameters we
can easily compare and find out how they vary from the Gaussian
distribution. It was seen that most of the cases were skewed
compared to the usual normal curve. Finally we have compared the
$f(R,T)$ models used in this study with the $\Lambda CDM$ model
and checked how much observational support the models enjoy. This
is done via a statistical mechanism, where AIC and BIC parameters
were calculated and compared. Then using the Jeffrey's scale we
could conclude which models are more efficient. It was seen that
the models with non-minimal coupling between matter and curvature
are observationally more favoured than the others. This idea which
is already present in the literature is re-confirmed from this
analysis. It must be mentioned here that none of the models used
in this analysis have $\Lambda$CDM cosmology as a limit. This is
quite clear from the considerable deviations of the AIC and BIC
values of the models in comparison to those of the $\Lambda$CDM
model. Moreover this is quite expected from the formulation of the
theory, where the matter sector have been coupled with the
curvature in the gravity Lagrangian itself. This creates extra
force giving rise to non-geodesic motion. The deviation from the
standard model is such that, it cannot be recovered as any
limiting case of these models. There may be questions regarding
the motivations of the models used in this work. So we would like
to mention here that these models are not cosmologically motivated
(except model V), and neither do they solve any particular issue.
But this work is important for the development of $f(R,T)$
theories, especially considering that fact that the models
considered here are generic in nature from the mathematical point
of view and can cover a family of models as limiting cases. So
here we are not solving any cosmological problem, but trying to
create viable theories of $f(R,T)$ with support from observational
data. These models being observationally constrained and favoured
may be used in future works on this theory for solving various
cosmological problems. So this work is a significant advancement
for $f(R,T)$ theory and modified gravity.

\section*{Acknowledgments}

PR acknowledges the Inter University Centre for Astronomy and
Astrophysics (IUCAA), Pune, India for granting visiting
associateship. KG acknowledges the High Performance Computing
System (HPC) at NITTTR Kolkata for using it as the computational
resource for this paper. Finally we thank the referee for his/her
invaluable comments that helped us to improve the quality of the
paper.


\section{Appendix : CC Data}
\begin{table}[h!]
\centering
\begin{tabular}{|c|c|c|c|c|c|}
\hline
  ~~~~~~$z$ ~~~~& ~~~~$H(z)$ ~~~~~& ~~~~$\sigma(z)$~~~~~& ~~~~$z$ ~~~~& ~~~~$H(z)$ ~~~~~& ~~~~$\sigma(z)$~~~~\\
  \hline
  0.07 & 69 & $\pm$ 19.6 & 0.4783 & 80.9 & $\pm$ 9\\
  0.09 & 69 & $\pm$ 12 & 0.48 & 97 & $\pm$ 62\\
  0.12 & 68.6 & $\pm$ 26.2 &  0.593 & 104 & $\pm$ 13\\
  0.17 & 83 & $\pm$ 8 & 0.68 & 92 & $\pm$ 8\\
  0.179 & 75 & $\pm$ 4 & 0.781 & 105 & $\pm$ 12\\
  0.199 & 75 & $\pm$ 5 & 0.875 & 125 & $\pm$ 17\\
  0.2 & 72.9 & $\pm$ 29.6 & 0.88 & 90 & $\pm$ 40\\
  0.27 & 77 & $\pm$ 14 & 0.9 & 117 & $\pm$ 23\\
  0.28 & 88.8 & $\pm$ 36.6 &  1.037 & 154 & $\pm$ 20\\
  0.352 & 83 & $\pm$ 14 & 1.3 & 168 & $\pm$ 17\\
  0.3802 & 83 & $\pm$ 13.5 & 1.363 & 160 & $\pm$ 33.6\\
  0.4 & 95 & $\pm$ 17 & 1.43 & 177 & $\pm$ 18\\
  0.4004 & 77 & $\pm$ 10.2 &  1.53 & 140 & $\pm$ 14\\
  0.4247 & 87.1 & $\pm$ 11.2 & 1.75 & 202 & $\pm$ 40\\
  0.44497 & 92.8 & $\pm$ 12.9 &  1.965 & 186.5 & $\pm$ 50.4\\ \hline
\end{tabular}

\caption{Cosmic Chronometer 30 point Data Set (Moresco, 2015). It
shows the 30 point cosmic chronometer $z-H(z)$ data with the
standard error $\sigma(z)$} \label{table:7}
\end{table}

\end{document}